\documentclass{article_saj}
\usepackage{graphicx,saj,multicol,subeqnarray}
\usepackage{natbib}
\usepackage{float}
\usepackage{xcolor}
\usepackage{widetext}
\usepackage{url}
\usepackage{bm}
\usepackage{tikz} 
\usepackage{pifont} 
\usepackage{amsmath,upgreek}
\definecolor{xlinkcolor}{cmyk}{1,0.6,0,0}
\usepackage[bookmarks=false,         
     pdfnewwindow=true,      
     colorlinks=true,    
     linkcolor=xlinkcolor,     
     citecolor=xlinkcolor,     
     filecolor=xlinkcolor,  
     urlcolor=xlinkcolor,      
final=true
]{hyperref}

\def\papertype{\ \hfill\ Editorial}

\def\udc{52}
\newcommand{\kepler}{{\it Kepler}}

\newcommand{\multi}{{\sc MultiNest}}

\def\checkmark{\tikz\fill[scale=0.4](0,.35) -- (.25,0) -- (1,.7) -- (.25,.15) -- cycle;}

\begin{document}
\parindent=.5cm
\baselineskip=3.1truemm
\columnsep=.5truecm
\newenvironment{lefteqnarray}{\arraycolsep=0pt\begin{eqnarray}}
{\end{eqnarray}\protect\aftergroup\ignorespaces}
\newenvironment{lefteqnarray*}{\arraycolsep=0pt\begin{eqnarray*}}
{\end{eqnarray*}\protect\aftergroup\ignorespaces}
\newenvironment{leftsubeqnarray}{\arraycolsep=0pt\begin{subeqnarray}}
{\end{subeqnarray}\protect\aftergroup\ignorespaces}
%


\markboth{\eightrm IMPOSSIBLE MOONS - TRANSIT TIMING EFFECTS THAT CANNOT BE DUE TO EXOMOON} 
{\eightrm KIPPING \& TEACHEY}

\begin{strip}

{\ }

\vskip-1cm

\publ

\type

{\ }


\title{IMPOSSIBLE MOONS - TRANSIT TIMING EFFECTS THAT CANNOT BE DUE TO EXOMOON}


\authors{D. Kipping$^{1,2}$ and A. Teachey$^1$}

\vskip3mm


\address{$^1$Department of Astronomy, Columbia University,
\break 550 W 120th Street, New York, NY 10027, USA}


\Email{dkipping@astro.columbia.edu}

\address{$^2$Center for Computational Astrophysics, Flatiron Institute,
\break 162 5th Av., New York, NY 10010, USA}



\dates{July 7, 2020}{October 13, 2020}


\summary{
Exomoons are predicted to produce transit timing variations (TTVs) upon their
host planet. Unfortunately, so are many other astrophysical phenomena - most
notably other planets in the system. In this work, an argument of
\textit{reductio ad absurdum} is invoked, by deriving the transit timing
effects that are impossible for a single exomoon to produce. Our work derives
three key analytic tests. First, one may exploit the fact that a TTV signal
from an exomoon should be accompanied by transit duration variations (TDVs),
and that one can derive a TDV floor as a minimum expected level of variability.
Cases for which the TDV upper limit is below this floor can thus be killed as
exomoon candidates. Second, formulae are provided for estimating whether moons
are expected to be ``killable'' when no TDVs presently exist, thus enabling the
community to estimate the value of deriving TDVs beforehand.
Third, a TTV ceiling is derived, above which exomoons should never be able to
produce TTV amplitudes. These tools are applied to a catalog of TTVs and TDVs for
two and half thousand \kepler\ Objects Interest, revealing over two hundred
cases that cannot be due to a moon - remarkably then a large fraction of the
known TTV amplitudes are consistent with being caused by a moon. These tests
are also applied to the exomoon candidate Kepler-1625b i, which comfortably
passes the criteria. These simple analytic results should provide a means of
rapidly rejecting putative exomoons and streamlining the search for satellites.
}



\end{strip}

\tenrm


\section{INTRODUCTION}

\indent

Transit timing variations (TTVs) have long been recognized as a powerful tool
for the detection of exoplanets \citep{dobro:1996a,dobro:1996b,miralda:2002,
schneider:2003,schneider:2004,holman:2005,agol:2005}, as well as exomoons
\citep{sartoretti:1999,szabo:2006,simon:2007,kipping:2009a,kipping:2009b}.
Early searches for TTVs, largely focussing on hot-Jupiters, were characterized
by either null detections \citep{steffen:2005,miller:2008a,miller:2008b,
rabus:2009,hrudkova:2010} or signals that later turned out to be spurious
\citep{diaz:2008,ribas:2009,maciejewski:2010} - so much so that one could be
forgiven for questioning the value of the TTV enterprise at the time.

This situation radically changed though with the launch of \kepler\
\citep{holman:2010,ballard:2011,nesvorny:2013,holczer:2016,hadden:2017} -
thanks to its detections of longer period planets and continuous,
long-baseline photometric observations. There are now hundreds of known
TTV systems, and this embarrassment of riches has actually presented a
problem. Specifically, TTVs due to planet-planet interactions are so
common that the search for exomoons is greatly frustrated by this enormous
background signal. In this era of abundant TTVs detections, there is a need
for tools that can quickly classify what TTVs can/cannot be - an era
of TTV triage.

In recent years, much of the theoretical work on TTVs has focussed on the
inverse problem \citep{nesvorny:2008,nesvorny:2010,lithwick:2012,nesvorny:2014,
deck:2016}. Yet these efforts broadly assume a specific model already - namely
that the observed TTVs are caused by another planet. A Bayesian would describe
this as parameter estimation. But parameter estimation is only one side of the
coin when it comes to inference, with the other being model selection.
Certainly, there are many cases where the model can be safely assumed to be
that of planet-planet interactions, for example because of the known existence
of near mean motion resonance transiting planets \citep{wu:2013,hadden:2014,
hadden:2017}. But it would be folly to assume that all TTVs will be universally
caused by planet-planet interactions - other models should be considered too.
And this is of course highly salient for exomoons, which represent a distinct
origin of TTVs.

Model selection for TTVs will be particularly challenging when the TTVs are
nearly sinusoidal, which is the form taken by circular orbit exomoons\footnote{
We highlight that this can be somewhat modified for very large moons that
distort the transit profile \citep{simon:2007}, but observational constraints on
the exomoon population of \kepler\ planets shows that this would be a rare
occurrence \citep{teachey:2017}.}
\citep{kipping:2009a}. This is because planet-planet interactions are perfectly
capable of appearing as sinusoidal, too, for example due to the circulating
line of conjunctions \citep{lithwick:2012,nesvorny:2014,deck:2016}. Thus, the
waveform shape of the TTVs will not necessarily be useful in distinguishing
these two hypotheses. In this work, using just the amplitude of the observed
transit timing effects, it is investigated whether this has any ability
to test the moon hypothesis. In particular, almost the opposite problem is
considered through an argument \textit{reductio ad absurdum} - what kind of
transit timing effects can one classify as being impossible for an exomoon?

\section{Killing Moons Below a TDV Floor}
\label{sec:TDVfloor}

\subsection{Conceptual explanation}

Before diving into the mathematical details of the effect described in this
section, it is instructive to first offer a simple intuitive explanation to
guide what follows. The TTV amplitude of an exomoon is proportional to the mass
of the satellite, $M_S$, multiplied by its semi-major axis, $a_S$
\citep{sartoretti:1999}. Consider that one has detected a TTV signal for an
exoplanet - a rather typical situation given that there are now hundreds of
such planets \citep{holczer:2016}. The task is now to determine if this data in
hand is consistent with an exomoon hypothesis, or not.

Consider that there exists an additional piece of information - measurements of
the TDVs. Exomoons are predicted to produce TDVs with the same periodicity as
the TTV signal, with the dominant TDV component being proportional to
$M_S a_S^{-1/2}$ \citep{kipping:2009a,kipping:2009b}. There are far fewer
examples of known TDV systems (see \citealt{szabo:2012} and
\citealt{nesvorny:2013} for rare examples) and so let's consider the more
typical case that a detected TTV signal does not appear to be accompanied by a
TDV signal.

This lack of a clear TDV signal can be translated into a TDV amplitude
upper limit, and that limit in fact places some interesting constraints on our
problem. For example, a moon in a compact orbit around its planet should
produce quite large TDVs, since the amplitude is proportional to $a_S^{-1/2}$.
Thus, one should expect that the lack of a TDV detection excludes these inner
orbits and places some limit on the minimum orbital radius of the hypothesized
exomoon. If the TDV upper limit is sufficiently tight, this minimum orbital
radius may in fact exceed the Hill sphere\footnote{One might question whether
quasi-moons outside of the Hill sphere defy this definition, but in this
work we consider that quasi-moon are exactly that - ``quasi-moons''
and not ``moons''.} - this would be an \textit{impossible moon}.

Thus, in what follows, the extent to which an exoplanet with a detected TTV
amplitude and a TDV upper limit can be used to deduce an minimum
allowed exomoon semi-major axis is investigated. If this semi-major axis
exceeds the Hill radius, then such cases could thus be dismissed as exomoon
candidates, despite the fact only a TTV signal was ever recovered.

\subsection{Mathematical details}

Let us begin by considering the ratio of the TDV amplitude to the TTV amplitude
as caused by an exomoon, denoted by the symbol $\eta$. This was first derived
in \citet{kipping:2009a}, who considered only velocity-induced TDVs (TDV-V).
The result is that $\eta = n_S T$, where $n_S$ is the mean motion of the moon
($=2\pi/P_S$) and $T$ is the planet's transit duration. The power of this
equation is that if both effects are detected, the orbital period of the moon
can be uniquely inferred, something usually not possible with exomoons due to
aliasing effects \citep{kipping:2009a}.

However, $\eta = n_S T$ is only true in the limiting case of i) no transit
impact parameter induced TDVs (TDV-TIP), caused by planets bobbing up and down
against the planet's orbital plane \citep{kipping:2009b} ii) zero eccentricity
for the satellite. Relaxing both of these assumptions, the ratio of the root
mean square (RMS) amplitudes is shown in \citet{thesis:2011} to be given by
(see their Equation~6.100), which is written here as

\begin{align}
\eta &= T n_S \underbrace{\frac{1}{(1-e_S^2)^{3/2}} \sqrt{\frac{\Phi_{\mathrm{TDV-V}}}{\Phi_{\mathrm{TTV}}}}}_{\equiv\zeta} + \epsilon,
\label{eqn:eta2}
\end{align}

where $\Phi_{\mathrm{TDV-V}}$ and $\Phi_{\mathrm{TTV}}$ are scalars controlling
the strength of the TDV-V and TTV effects respectively (which depend on the
three-dimensional geometry of the orbits), $\epsilon$ is a
parameter introduced here to absorb the TDV-TIP component of the $\eta$
parameter (see Equation~6.100 of \citealt{thesis:2011} for full form).

Given that the above is a generalization to eccentric satellite orbits, it
is not surprising that the $\zeta$ term defined in Equation~(\ref{eqn:eta2})
tends to unity in the limit of $e_S \to 0$, which is evident from the
behaviour of the $(1-e_S^2)^{-3/2}$ term, as well as the fact that

\begin{align}
\lim_{e_S \to 0} \Phi_{\mathrm{TTV}} = \lim_{e_S \to 0} \Phi_{\mathrm{TDV-V}} = \pi (1 - \cos^2i_S\sin^2\varpi_S),
\label{eqn:Philimit}
\end{align}

as shown in Equations~(6.47) \& (6.67) of \citet{thesis:2011}. Although $\zeta$
tends towards unity for circular moons, one might wonder whether it is
typically greater than unity (i.e. an enhancement factor to $\eta)$ or less
than unity (i.e. an attenuation factor) for $e_S>0$. Since the functional
forms of the $\Phi_{\mathrm{TTV}}$ and $\Phi_{\mathrm{TDV-V}}$ are rather
protracted, it is not straight-forward to analytically investigate this
behaviour. Instead, a large number ($10^7$) of random examples of
$0\leq e_S<1$, $0\leq\omega_S<2\pi$, $0\leq\varpi_S<2\pi$ and $0\leq i_S<2\pi$
were generated, with a corresponding calculation of $\zeta$. From this, it is
found that $\zeta \geq 1$ in every single numerical test, showing that it can
only ever serve as an enhancement factor. Generally, one does not expect moons
to possess large eccentricities due to tidal circularization, but the following
argument holds even in such a case.

Next, consider the behaviour of $\epsilon$, which relates to the TDV-TIP
effect. The TDV-TIP effect is offset in phase from the TDV-V effet by
$\pm\pi/2$ radians (depending on whether the moon is prograde or retrograde;
see \citealt{kipping:2009b}). For this reason, it can never destructively
interfere with the TDV-V effect to attenuate the signal. Rather, whatever
the amplitude of the TDV-TIP effect, and whether it be prograde or retrograde,
it can only act to increase the overall TDV amplitude. Thus, $\epsilon>0$ in
all cases. Therefore, like the $\zeta$ term, $\epsilon$ can only act as an
enhancement factor to $\eta$. Generally, the TDV-TIP effect is small compared
to the TDV-V effect \citep{kipping:2009b}, but this is not actually a
requisite in the following argument. To summarize the results so far, it has
been shown that

\begin{align}
\eta &= \frac{\delta_{\mathrm{TDV}}}{\delta_{\mathrm{TTV}}} = \zeta T n_S + \epsilon,
\end{align}

where $\delta_{\mathrm{TTV}}$ and $\delta_{\mathrm{TDV}}$ are TTV and TDV RMS
amplitudes (respectively), $T$ is the transit duration, $n_S$ is the
satellite's mean motion, and $\zeta$ and $\epsilon$ are scalars such that
$\zeta\geq1$ and $\epsilon>0$.

With these points established, one may now consider how $\eta$ can be
used to identify impossible moons. Let us replace $P_S$ with $P_P$ using
Equation~(12) of \citet{kipping:2009a}, which establishes that $P_S \simeq 
P_P \sqrt{f^3/3}$, yielding

\begin{align}
\frac{\delta_{\mathrm{TDV}}}{\delta_{\mathrm{TTV}}} &= \zeta \Bigg(\frac{2\pi \sqrt{3} T}{P_P f^{3/2}}\Bigg)  + \epsilon,
\end{align}

where $f$ is the semi-major axis of the moon relative to the Hill radius of the
planet (and thus one expects $f<1$). Consider that one has an upper limit on
$\delta_{\mathrm{TDV}}$ given by $\delta_{\mathrm{TDV,max}}$, such that
$\delta_{\mathrm{TDV}}\leq\delta_{\mathrm{TDV,max}}$. Substituting this
into our $\eta$ relation gives

\begin{align}
\frac{\delta_{\mathrm{TDV,max}}}{\delta_{\mathrm{TTV}}} &\geq \zeta \Bigg(\frac{2\pi \sqrt{3} T}{P_P f^{3/2}}\Bigg) + \epsilon.
\end{align}

Since the left hand side (LHS) of the equation is always greater than the
right hand side (RHS), then the
inequality will also be true in the case of $\epsilon \to 0$ and
$\zeta \to 1$, since these limits represent the smallest allowed values
for these terms. Accordingly, one may write - without any loss of generality 
- that

\begin{align}
\frac{\delta_{\mathrm{TDV,max}}}{\delta_{\mathrm{TTV}}} &\geq \frac{2\pi \sqrt{3} T}{P_P f^{3/2}}.
\end{align}

In the above, essentially all terms are observable under the stated assumptions
of the problem - except for $f$. One may thus rearrange to make $f$ the
subject:

\begin{align}
f^{3/2} &\geq \Bigg(\frac{2\pi \sqrt{3}T}{P_P}\Bigg) \Bigg(\frac{\delta_{\mathrm{TTV}}}{\delta_{\mathrm{TDV,max}}}\Bigg).
\end{align}

Finally, one can see that the above represents a lower limit on $f$,
denoted as $f_{\mathrm{min}}$, and given by

\begin{align}
f_{\mathrm{min}} &= \Big( \frac{2\pi \sqrt{3}T}{P_P} \Big)^{2/3} \Big(\frac{\delta_{\mathrm{TTV}}}{\delta_{\mathrm{TDV,max}}}\Big)^{2/3}.
\label{eqn:fmin}
\end{align}

\begin{figure*}
\begin{center}
\includegraphics[width=17.4cm]{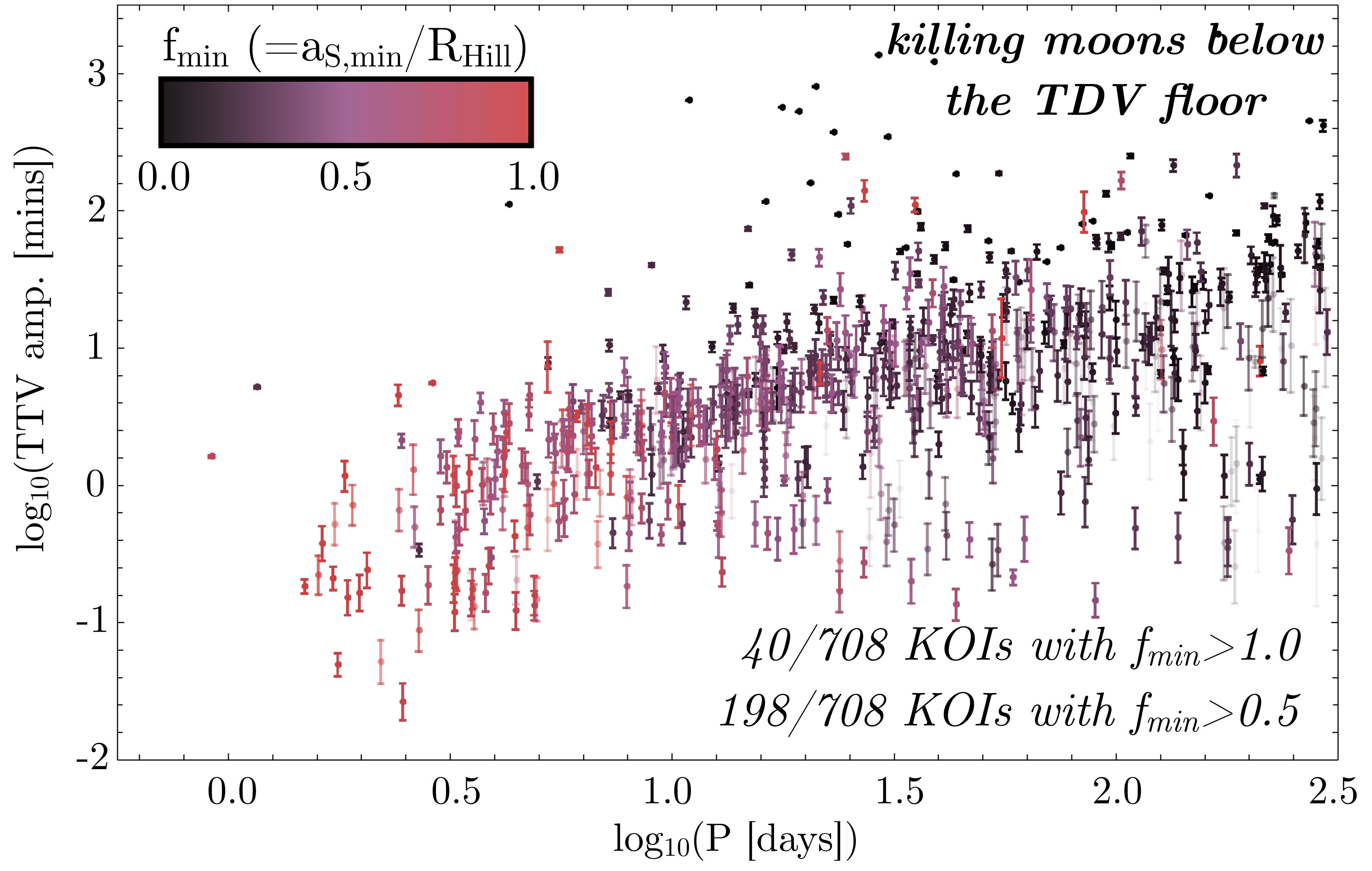}
\caption{
Best-fitting TTV amplitude from a LS periodogram applied to the
outlier-cleaned \citet{holczer:2016} TTV catalog, as a function of planetary
period. Point opacity equals the $\Delta BIC/10$ score of the fits. The
color coding shows the minimum allowed exomoon semi-major axis (given by
Equation~\ref{eqn:fmin}), assuming that the TTVs are caused by a single
satellite and conditioned upon the fact no TDVs are observed. 40 KOIs are
identified that require $f>1$ and can thus be killed as exomoon candidates.
}
\label{fig:tdv}
\end{center}
\end{figure*}

Figure~\ref{fig:tdv} shows some example calculations of $f_{\mathrm{min}}$ for
an ensemble of KOIs (\kepler\ Objects of Interest) with available TDVs (taken
from \citealt{holczer:2016}), with the data methods described later in
Section~\ref{sec:applied}.

Since a moon should always have $f<1$ \citep{domingos:2006}, one can write
a simple criterion that a real exomoon should satisfy:

\begin{align}
\Big( \frac{2\pi \sqrt{3}T}{P_P} \Big)^{2/3} \Big(\frac{\delta_{\mathrm{TTV}}}{\delta_{\mathrm{TDV,max}}}\Big)^{2/3} < 1,
\end{align}

or more succintly, one can define a ``TDV floor'' criterion is

\begin{align}
\delta_{\mathrm{TDV,max}} > 2\pi \sqrt{3} \Bigg( \frac{T \delta_{\mathrm{TTV}}}{P_P} \Bigg).
\label{eqn:TDVfloor}
\end{align}

To illustrate this, consider the case where the TDV limits are very noisy, such
that $\delta_{\mathrm{TDV,max}}$ is poorly constrained, with a very large upper
limit. In such a case, the criterion is satisfied. This does not prove the
solution is an exomoon, but it means that the current observational constraints
are at least consistent with said hypothesis. Now imagine observers obtaining
ever more precise TDVs, yet still no significant detection ever emerges, thus
gradually lowering $\delta_{\mathrm{TDV,max}}$. Eventually, the criterion
will fail and at that point one can confidently assert that the observations
are inconsistent with being caused by a single large exomoon - an impossible
moon.

\section{The Killing Regime}
\label{sec:killable}

If TDV measurements are expected to be very noisy or too few in number, one
might be able to immediately conclude that it's not even worth the effort of
attempting to derive TDVs to test the criterion of
Equation~(\ref{eqn:TDVfloor}). The fundamental noise limit itself will
not be sensitive enough to infer an impossible moon. This is certainly a
worthwhile point to consider because TDVs can be computationally expensive
to derive. In what follows then, the expected upper limit on the TDV
amplitude is derived under some simplifying assumptions.

Consider a sequence of $N$ homoscedastic TDVs with normally distributed noise
of standard deviation $\Delta_{\mathrm{TDV}}$. Generally, moons are expected
to be near circular, producing sinusoidal TDVs \citet{kipping:2009a}. It is
here assumed that the TDVs are dominated by a single component to simplify the
analysis, which will typically be the TDV-V effect (except for grazing
transiting planets or highly inclined moons; \citealt{kipping:2009b}).
Accordingly, the model being regressed to the data is

\begin{align}
\mathrm{TDV}(E) &= A_{\mathrm{TDV}} \sin(n E + \phi),
\end{align}

where $E$ is the epoch number, $\phi$ is a phase term and $n$ is the TDV
frequency. The killing criterion described in Section~\ref{sec:TDVfloor} is
for the case where a strong TTV has already been detected. If the signal is
due to an exomoon, then the TDVs will have the same periodicity as the TTVs,
and that period is presumably well-constrained thanks to the TTV detection.
Further, the phase is also known, since TDV-Vs lag TTVs by $\pi/2$ radians
\citep{kipping:2009a}. Accordingly, the only free parameter is the amplitude,
which is closely related to the RMS amplitude via
$A_{\mathrm{TDV}}=\sqrt{2} \delta_{\mathrm{TDV}}$.

Since the data are assumed to be normally distributed, and the problem is
linear with respect to the one unknown parameter, then one may employ linear
least squares regression theory to write that the parameter covariance matrix
(a one-by-one matrix in this case) will be given by

\begin{align}
\bm{\Sigma} &= \Delta_{\mathrm{TDV}}^2 (\mathbf{X}^T \mathbf{X})^{-1},
\end{align}

where the homoscedasticity of the problem has been exploited, and $\mathbf{X}$
is given by

\begin{eqnarray}
\mathbf{X} =
\begin{bmatrix}
	\sin(n E_1 + \phi)\\
	\sin(n E_2 + \phi)\\
	...\\
	\sin(n E_N + \phi)
\end{bmatrix}.
\end{eqnarray}

Evaluating, one may show that

\begin{align}
\bm{\Sigma} &= \frac{ \Delta_{\mathrm{TDV}}^2 }{ \sum_{i=1}^N \sin^2(n E_i + \phi) },
\end{align}

and thus the error on $A_{\mathrm{TDV}}$ will be

\begin{align}
\sigma_{\mathrm{A}_{\mathrm{TDV}}} = \frac{ \Delta_{\mathrm{TDV}} }{ \sqrt{\sum_{i=1}^N \sin^2(n E_i + \phi)} }.
\end{align}

Sampling of the TDV curve is random, there is no preference for any particular
phase to be observed. This means that $\bm{\Sigma}$ will not always
return the same covariance matrix even for the same number of points with the
same uncertainty - the term is probabilistic. It is therefore necessary to
estimate the expectation value of $\bm{\Sigma}$ accounting for this
feature.

One can write that $\sin^2(n E_i + \phi) \to \sin^2(x_i)$, where $x_i$ is a
uniform random variate between 0 and $2\pi$. The probability distribution of
$\sin^2x_i$ is now well-posed, and described by the arc-sine distribution such
that

\begin{align}
\mathrm{Pr}(y_i=\sin^2x_i)\,\mathrm{d}y_i &= \frac{1}{\pi \sqrt{y_i} \sqrt{1-y_i}}\,\mathrm{d}y_i.
\end{align}

The expectation value of the TDV uncertainty now becomes

\begin{align}
\frac{E[\sigma_{\mathrm{A}_{\mathrm{TDV}}}]}{\Delta_{\mathrm{TDV}}} = E\Big[ \frac{1}{\sqrt{\sum_{i=1}^N y_i}} \Big].
\end{align}

In the case of $N\gg1$, the expectation value of the RHS becomes $2/\sqrt{N}$,
such that

\begin{align}
\lim_{N\gg1} E[\sigma_{\mathrm{A}_{\mathrm{TDV}}}] &= \frac{2\Delta_{\mathrm{TDV}}}{\sqrt{N}},\nonumber\\
\lim_{N\gg1} E[\sigma_{\delta_{\mathrm{TDV}}}] &= \frac{\sqrt{2}\Delta_{\mathrm{TDV}}}{\sqrt{N}}.
\label{eqn:noisy}
\end{align}

The upper limit on the TDV amplitude can be expressed as some factor of this
noise estimate, with a typical choice being 3. Accordingly, it is estimated that
a null TDV signal will have an upper limit of 

\begin{align}
\delta_{\mathrm{TDV,max}} &\simeq 3 \times \frac{\Delta_{\mathrm{TTV}}}{\sqrt{2N}},
\label{eqn:TDVmaxpred}
\end{align}

where the replacement $\Delta_{\mathrm{TDV}} \simeq 2\Delta_{\mathrm{TTV}}$ has
also bee used - i.e. the duration error is approximately twice the timing error
\citep{carter:2008}. This replacement is necessary since the assumption
throughout is that the TDVs have not yet been derived and one is deciding as to
whether it's worth inferring them. Combining this with our earlier
Equation~(\ref{eqn:TDVfloor}) yields the following requirement for an physical
exomoon - assuming a TTV signal has been detected and a TDV upper limit:




\begin{align}
\Bigg( \frac{\delta_{\mathrm{TTV}}}{\sqrt{2} \Delta_{\mathrm{TTV}}/\sqrt{N}} \Bigg) < \Bigg( \frac{\sqrt{3}}{4\pi} \Bigg) \Bigg(\frac{P_P}{T}\Bigg).
\end{align}

Although the above is more accurate, it is useful to convert it into a more
intuitive form by replacing the denominator on the LHS with
$\sigma_{\delta_{\mathrm{TTV}}}$ via analogy to Equation~(\ref{eqn:noisy}).
This is somewhat inaccurate because the TTV fit was not a one-parameter fit,
and so the real uncertainty may be greater than this due to parameter
covariances. Accordingly, the above form is recommended but for the sake
of guiding intuition, one may write that


\begin{align}
\Bigg( \frac{\delta_{\mathrm{TTV}}}{\sigma_{\delta_{\mathrm{TTV}}}} \Bigg) < \Bigg( \frac{\sqrt{3}}{4\pi} \Bigg) \Bigg(\frac{P_P}{T}\Bigg).
\label{eqn:noisycriterion}
\end{align}

With the derivation complete, let us take a step back and interpret what
has actually been derived. Recall that the inequality imposed is the condition
for a plausible moon stemming from Equation~(\ref{eqn:TDVfloor}). How should
one interpret the above? Note that the LHS of
Equation~(\ref{eqn:noisycriterion}) now represents the TTV signal-to-noise
ratio. Therefore, if you have TTV detection with a signal-to-noise of $>
P_P/(7T)$, then one should expect that derived TDVs will be capable of
killing the exomoon hypothesis (assuming they don't see anything). In such
cases, if the moon hypothesis is considered viable, it would instructive to
derive TDVs then, since their absence would falsify the moon hypothesis.

\begin{figure*}
\begin{center}
\includegraphics[width=17.4cm]{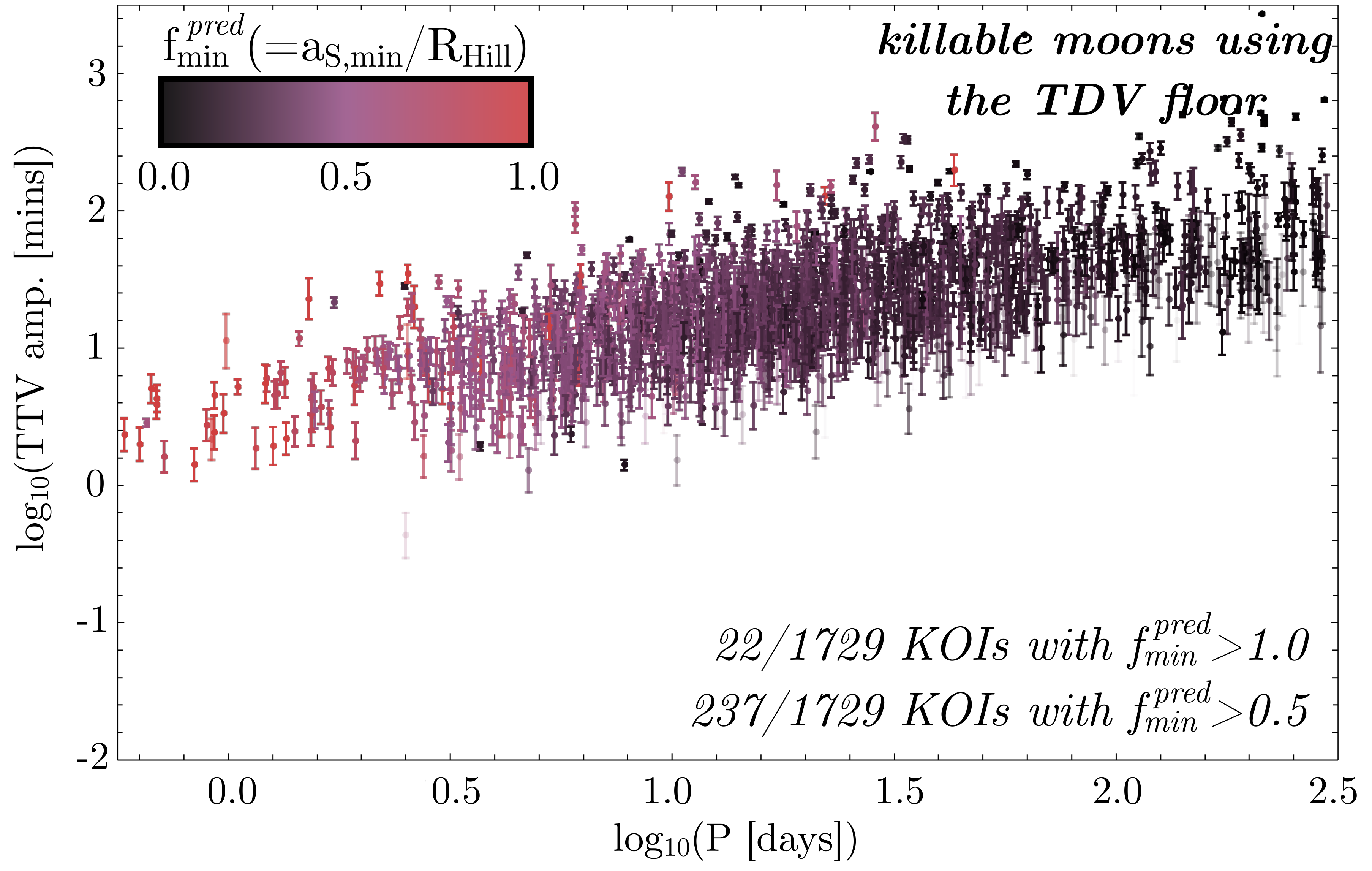}
\caption{
Same as Figure~\ref{fig:tdv} except showing only KOIs for which no TDVs are
available. Here, it is assumed that no TDVs will be found and so what is
shown is a predicted limit on $f$ using the expected TDV sensitivity from
Equation~(\ref{eqn:TDVmaxpred}) and the minimum $f$ expression of
Equation~(\ref{eqn:fmin}). Almost two dozen KOIs with large TTVs are identified
that should be ``killable'' as exomoons, if TDVs were to become available.
}
\label{fig:tdv_pred}
\end{center}
\end{figure*}

What if Equation~(\ref{eqn:noisycriterion}) is not satisfied? In such case,
TDVs will be less valueable when it comes to seeking exomoons, and thus
might be lower priority objects - but even here TDVs will still have some
utility. Notably, although one cannot guarantee that the TDVs
will be capable of excluding the moon hypothesis, they will still
place important constraints. For example, they will still place a minimum
constrain on $f$ via Equation~(\ref{eqn:fmin}) - albeit a minimum $f$ which
lies within the Hill sphere. This is illustrated in Figure~\ref{fig:tdv_pred}
for example (with more specific details about the data provided later in
Section~\ref{sec:applied}). This $f$ constraint may still be sufficient to
place tension on the moon hypothesis, since only retrograde moons are thought
to be dynamically stable beyond $f>0.5$ for example
\citep{domingos:2006}. Alternatively, they may actually lead to a TDV
detection, thus lending support to the moon hypothesis.

\section{The TTV Ceiling}
\label{sec:ttvceiling}

%
In the previous section, the focus was on the ratio of the TTV amplitude to the
upper limit for the TDVs. In some cases, TDV upper limits are not presently
available and so it is useful to consider if there is some maximum to
how strong the TTV effect of an exomoon can be, in an absolute sense - a 
TTV ceiling. Specifically, moons cannot generate
TTVs of arbitrarily large amplitude and so TTVs above this ceiling can
be immediately flagged as unrealistic to be caused by a moon.

The TTV amplitude of an exomoon was first derived in \citet{sartoretti:1999}
for the case of circular, coplanar orbits. This calculation was generalized
to arbitrary orbits in \citet{kipping:2009a}, who found that the RMS amplitude
is given by

\begin{align}
\delta_{\mathrm{TTV}} =&
\Bigg(\frac{1}{2\pi}\Bigg)
\Bigg(\frac{a_S M_S P_P}{a_P M_P}\Bigg)
\Bigg(\frac{ (1-e_S)^2 \sqrt{1-e_P^2} }{ 1+e_P\sin\omega_P}\Bigg)\nonumber\\
\qquad=& \times
\Bigg(\sqrt{ \frac{\Phi_{\mathrm{TTV}}}{2\pi} }\Bigg).
\end{align}

It is first highlighted that the planetary eccentricity terms can be replaced
with stellar density observables via the photoeccentric effect
\citep{dawson:2012,MAP:2012}, such that:

\begin{align}
\Psi^{1/3} = \frac{ 1 + e_P \sin\omega_P }{\sqrt{1-e_P^2}} = \Bigg(\frac{\rho_{\star,\mathrm{obs}}}{\rho_{\star}}\Bigg)^{1/3},
\end{align}

where $\rho_{\star}$ is the mean density of the host star and
$\rho_{\star,\mathrm{obs}}$ is the value inferred from a circular orbit fit to
the transit light curve. Substituting this into the TTV equation gives

\begin{align}
\delta_{\mathrm{TTV}} &=
\Bigg(\frac{1}{2\pi}\Bigg)
\Bigg(\frac{a_S M_S P_P}{a_P M_P}\Bigg)
\Bigg(\frac{ (1-e_S)^2 }{ \Psi^{1/3} } \Bigg)
\Bigg(\sqrt{ \frac{\Phi_{\mathrm{TTV}}}{2\pi} }\Bigg).
\end{align}

Next, let us use substitute $a_S = f R_{\mathrm{Hill}}$, where
$R_{\mathrm{Hill}} \equiv a_P \sqrt[3]{M_P/(3M_{\star})}$, to give

\begin{align}
\delta_{\mathrm{TTV}} =&
\Bigg(\frac{1}{2\pi}\Bigg)
\Bigg(\frac{f M_S P_P}{M_P}\Bigg)
\Bigg(\frac{M_P}{3M_{\star}}\Bigg)^{1/3}
\Bigg(\frac{ (1-e_S)^2 }{ \Psi^{1/3} } \Bigg)\nonumber\\
\qquad& \times \Bigg(\sqrt{ \frac{\Phi_{\mathrm{TTV}}}{2\pi} }\Bigg).
\end{align}

Let's now proceed to maximize the RHS in order to derive a ceiling for the TTV
amplitude. By definition, a moon must satisfy $M_S \leq M_P$ and using this
in the above yields the inequality:

\begin{align}
\delta_{\mathrm{TTV}} &\leq f
\Bigg(\frac{\Psi^{-1/3}}{n_P}\Bigg)
\Bigg(\frac{q}{3}\Bigg)^{1/3}
\underbrace{\Bigg((1-e_S)^2 \sqrt{ \frac{\Phi_{\mathrm{TTV}}}{2\pi} }\Bigg)}_{\equiv\beta},
\label{eqn:temp1}
\end{align}

where $q \equiv (M_P/M_{\star})$ and a new term, $\beta$, is defined. It is
noted that one could also compute the above with some smaller choice of $M_S$
besides from the the binary limit assumed here, such as one based on a system
age plus tidal migration \citep{barnes:2002}. In what follows though, it is
preferred to keep the limit as broad as possible to avoid erroneously removing
massive moons with unanticipated origins/evolution.

The eccentricity of the satellite is unknown and resides somewhere in the range
$0\leq e_S < 1$ for a stable satellite. Consider the limiting case of
$e_S \to 0$, where \citet{thesis:2011} shows (Equation~6.47) that

\begin{align}
\lim_{e_S \to 0} \Phi_{\mathrm{TTV}} = \pi (1- \cos^2i_S \sin^2\varpi_S)
\end{align}

The cosine and sine squared terms must always be in the range
of zero to unity, and thus

\begin{align}
0 \leq \Big( \lim_{e_S \to 0} \Phi_{\mathrm{TTV}}\Big) \leq \pi
\end{align}

which means that

\begin{align}
0 \leq \lim_{e_S\to0} \beta \leq \frac{1}{\sqrt{2}}.
\end{align}

Thus, in the limit of circular moons, this term can only ever be smaller than
$1/\sqrt{2}$. If one wishes to maximize the RHS of Equation~(\ref{eqn:temp1})
then, one may simply set this combined term to that value. However, this is
only true for $e_S \to 0$ and so let us now consider what the effect of moon
eccentricity would be. The $(1-e_S^2)$ term in front rapidly drops to zero and
outpaces the divergent behavior of $\Phi_{\mathrm{TTV}}$, causing the combined
function to tend to zero as $e_S\to1$, which is of course less than
$1/\sqrt{2}$. To consider intermediate eccentricities, between 0 and 1, $10^7$
Monte Carlo samples were again generated as was done in
Section~\ref{sec:TDVfloor} earlier. From this, it was found that the maximum
occurs close to (but not exactly) $i_S \to \pi/2$, $\omega_S \to \pi/2$ and
$\varpi_S \to 3\pi/2$, for which

\begin{align}
\lim_{i_S\to\pi2/} \lim_{\omega_S\to\pi/2} \lim_{\varpi_S\to3\pi/2} \beta = \sqrt{ \frac{
1 + e_S^2 (\sqrt{1-e_S^2}-1)}{
1 + \sqrt{1-e_S^2}
}}.
\end{align}

The RHS is maximized when $e_S = 0.57747...$ for which it evaluates to
$\beta=0.71891$. Since this isn't quite the exact maximum, it was compared to
the numerical search which finds the largest ever value of $\beta$ was
$0.719734$. Thus, even with very careful fine tuning, the $\beta$
term isn't able to noticeably rise above $1/\sqrt{2}=0.707...$ in value.
Setting this as a limit then, Equation~(\ref{eqn:temp1}) becomes

\begin{align}
\delta_{\mathrm{TTV}} &\leq f
\Bigg(\frac{\Psi^{-1/3}}{\sqrt{2} n_P}\Bigg)
\Bigg(\frac{q}{3}\Bigg)^{1/3}
\end{align}

It is now useful to write the amplitude in units of the planetary period,
yielding a fractional TTV, and to move the $\sqrt{2}$ next to the RMS
amplitude such that it equates to a sinusoid amplitude, yielding

\begin{align}
\Bigg(\frac{\sqrt{2}\delta_{\mathrm{TTV}}}{P_P}\Bigg) &\leq
\frac{f}{9}
\Bigg(\frac{q}{\Psi}\Bigg)^{1/3}.
\end{align}

For planets with $\Psi$ deviating from unity by a large amount, the system
would probably be considered suspicious as an exomoon host on the basis that
the planet have a high eccentricity \citep{domingos:2006}. Thus, generally, one
expects $\Psi \sim 1$, yielding

\begin{align}
\lim_{e_P \ll 1} \Bigg(\frac{A_{\mathrm{TTV}}}{P_P}\Bigg) &<
\frac{f q^{1/3}}{9}.
\label{eqn:ttvceiling}
\end{align}

The above expression represents a TTV ceiling for exomoons, where a
reasonable choice for $f$ would be unity. However, since the TTV
amplitude is an observable, one can also rearrange the above to give
another lower limit on $f$:

\begin{align}
\lim_{e_P \ll 1} f_{\mathrm{min}} = \Bigg(\frac{9}{q^{1/3}}\Bigg) \Bigg(\frac{A_{\mathrm{TTV}}}{P_P}\Bigg)
\label{eqn:fmin2}
\end{align}

\begin{figure*}
\begin{center}
\includegraphics[width=17.4cm]{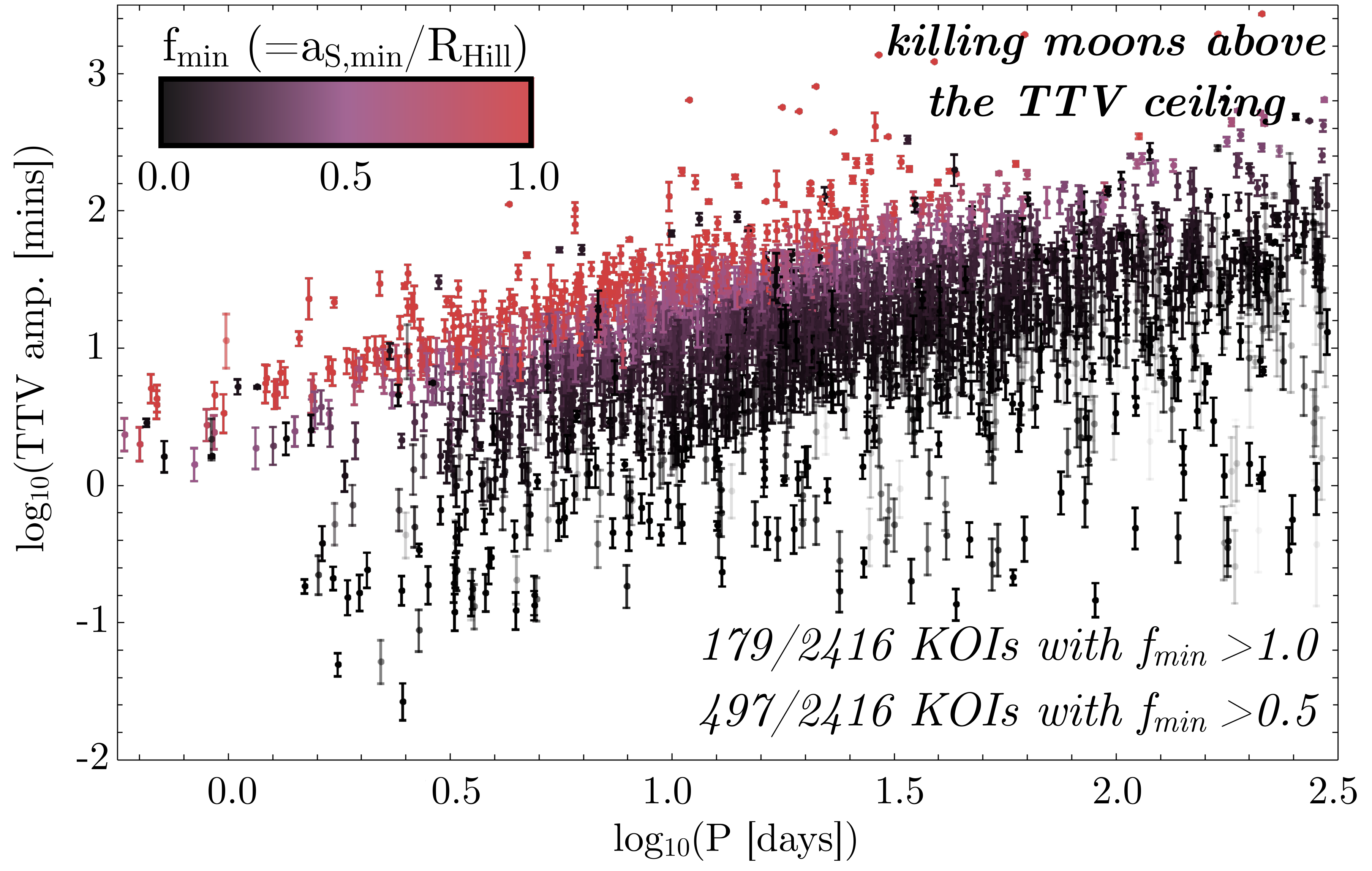}
\caption{
Same as Figure~\ref{fig:tdv} except the minimum exomoon semi-major
axis (normalized by the planetary Hill radius) is calculated using
the TTV ceiling effect of Equation~(\ref{eqn:fmin2}). On this basis,
179 KOIs can be rejected as being plausibly due to a single
exomoon, shown in red. Note that all 2437 KOIs are not shown from the full
sample, since 21 do not have planetary radii necessary to compute the
TTV ceiling and are thus removed.
}
\label{fig:ttv}
\end{center}
\end{figure*}

If an estimate of $q$ is available, it is therefore straight-forward to
evaluate either of these equivalent expressions. Since $q$ is often not
known for transiting planets, it is recommend here that one use an upper
limit for the $q$ estimate, so that the above is a conservative evaluation.
This is done for the \citet{holczer:2016} KOI sample in Figure~\ref{fig:ttv},
with more details provided later in Section~\ref{sec:applied}.

As a more general example, the Jupiter-Sun pair has $q \sim 10^{-3}$ and thus
one would not expect fractional TTV amplitudes greater than 1\%. For smaller
planets, such as the Earth-Sun pair, the maximum allowed fractional TTV
amplitude  drops another order-of-magnitude to 0.1\%. This provide some crude
cuts for removing suspiciously large TTV amplitudes in the search for moons.

One might wonder if a TDV ceiling also exists. Recall that the TDVs from an
exomoon have two quite distinct components, TDV-V and TDV-TIP. The TDV-V
effect is maximized as the moon's semi-major axis tends to zero. Of course,
in practice this cannot happen since the moon would impact the planet.
However, existing TDV-V theory does not extend to highly compact moons
since the derivation of \citet{kipping:2009a,kipping:2009b,thesis:2011}
explicitly assume that the moon's velocity is constant during the transit
duration. Once the moon's period becomes comparable to the duration or less,
this is no longer true.

Further, it is highlighted that the TDV-TIP does not have any true upper limit
since it can bob a planet in and out of transit entirely, in principle.
Thus, one would obtain a series of missed transits within the overall
sequence. For these reasons, combined with the fact any kind of TDV
detection is relatively rare, no effort was made to deduce a TDV ceiling
here.

\section{Application}
\label{sec:applied}

\subsection{Applying to the Holczer et al. (2016) Catalog}

To apply these formulae, one requires a homogeneous catalog of both TTVs and
TDVs for a sample of transiting planets. To this end, the \citet{holczer:2016}
catalog is utilized, which includes TTVs for 2599 \kepler\ Objects of Interest
(KOIs). It was decided to perform an independent analysis of these TTVs, in
order to measure putative TTV amplitudes for each, the statistical significance
of said signals, and upper limits on the TDVs. The catalog includes TTVs for
all 2599 KOIs and TDVs for cases where the authors deemed the data quality was
sufficient to attempt their derivation.

First, for each KOI, the dispersion of the TTVs is recorded by measuring
the scatter of the TTV measurements divided by their uncertainties. This
essentially tracks how many ``sigmas'' the TTV points are dispersed about zero.
To account for possible outliers, the median deviation multiplied by 1.4286 is
used as our measure of scatter. Any TTV measurements which exhibit a deviation
from zero greater than 10 times this value are then removed. A further cull is
applied to any TTV measurements where the TTV uncertainty is more than 3 times
greater than the median TTV uncertainty. The same process is applied to the
TDVs, where available.

Next, a Lomb-Scargle (LS) periodogram \citep{lomb:1976,scargle:1982} is run
through the TTVs using a log-uniformly spaced grid of periods from the Nyquist
period out to twice the baseline of observations. The log-period spacing of
this grid was set to $\log_2(0.01)$. At each period, the best fitting sinusoid
is computed using a weighted linear least squares regression, saving the
$\chi^2$ of the fit and the amplitude. Once the periodogram is complete, the
peak of highest $\chi^2$ improvement over a flat line is saved, whose period,
amplitude and $\chi^2$ are used in what follows. These TTV amplitudes comprise
the $y$-axis information of Figures~\ref{fig:tdv}, \ref{fig:tdv_pred} \&
\ref{fig:ttv}, where the TTV amplitude uncertainty is assigned using
Equation~(\ref{eqn:noisy}) - but with the ``TDV'' subscripts replaced with
``TTV''.

Next, a filter is applied to only accept KOIs for which the best sinusoidal fit
is statistically favoured over a simple flat line. This is accomplished by
calculating the Bayesian Information Criterion (BIC) difference between the two
models \citep{schwarz:1978}, and selecting only KOIs where the sinusoid yields
an improved BIC. From this, it was found that the vast majority, 2437 of the
2599 KOIs, favor the sinusoidal TTV model. Indeed, 1810 of these would be classed
as ``very strong'' ($\Delta\mathrm{BIC}>10$) using the \citet{kass:1995}
interpretative scale. This highlights just how valuable tools to quickly
classify this large number of detections can be. The TTV amplitudes and
$\Delta\mathrm{BIC}$ scores for each of 2437 KOIs are listed in
Table~\ref{tab}.

For the TDVs, one can first exploit the fact that the hypothesis that is being
tested requires the TDV period to be equal to the TTV period
\citep{kipping:2009a,kipping:2009b}. One can therefore simply fix the TDV
period to that resulting from the earlier TTV periodogram. It also noted TDVs
due to an exomoon are generally expected to be dominated by TDV-Vs
\citep{kipping:2009b} with a phase-lag of $\pi/2$ radians. This reduces the TDV
fit to a single parameter; amplitude. This also makes the estimate of TDV error
and 3\,$\sigma$ upper limit straight-forward, since for a one-parameter model 
such as this one can simply use $\Delta\chi^2$ to extract errors.

After running through all KOIs, only two KOIs were found with a strong preference
for a coupled TDV signal ($\Delta\mathrm{BIC}>10$), KOI-1546.01 and
KOI-5802.01, which may deserve further attention. For the others, the TDV
3\,$\sigma$ upper limit is used to compute $f_{\mathrm{min}}$ via
Equation~(\ref{eqn:fmin}), where basic transit parameters (e.g. transit
duration) are taken from the NASA Exoplanet Archive \citep{akeson:2013}. From
this, Figure~\ref{fig:tdv} is produced, which identifies 40 ``impossible
moons'' from 708 KOIs with TDVs. If the constraint is tightened such that only
prograde moons are permitted ($f<0.5$), then this substantially increases to
198 KOIs.

In cases without any available TDVs, it is instead investigated what constraint
on $f_{\mathrm{min}}$ one might expect them to provide, if they were to be
derived. This is done following the methods described in
Section~\ref{sec:killable}, where it is assumed that no TDVs will be found
and calculate a predicted limit on $f$ using expected TDV sensitivity from
Equation~(\ref{eqn:TDVmaxpred}) and the minimum $f$ expression of
Equation~(\ref{eqn:fmin}). Figure~\ref{fig:tdv_pred} shows the result of
this exercise, where identify 20 KOIs (out of 1729 without TDVs) for which
TDVs measurements should be able to completely exclude the exomoon hypothesis,
assuming no TDVs are found. Again, this steeply rises to 237 if one allows
prograde moons only.

\begin{figure}
\begin{center}
\includegraphics[width=8.4cm]{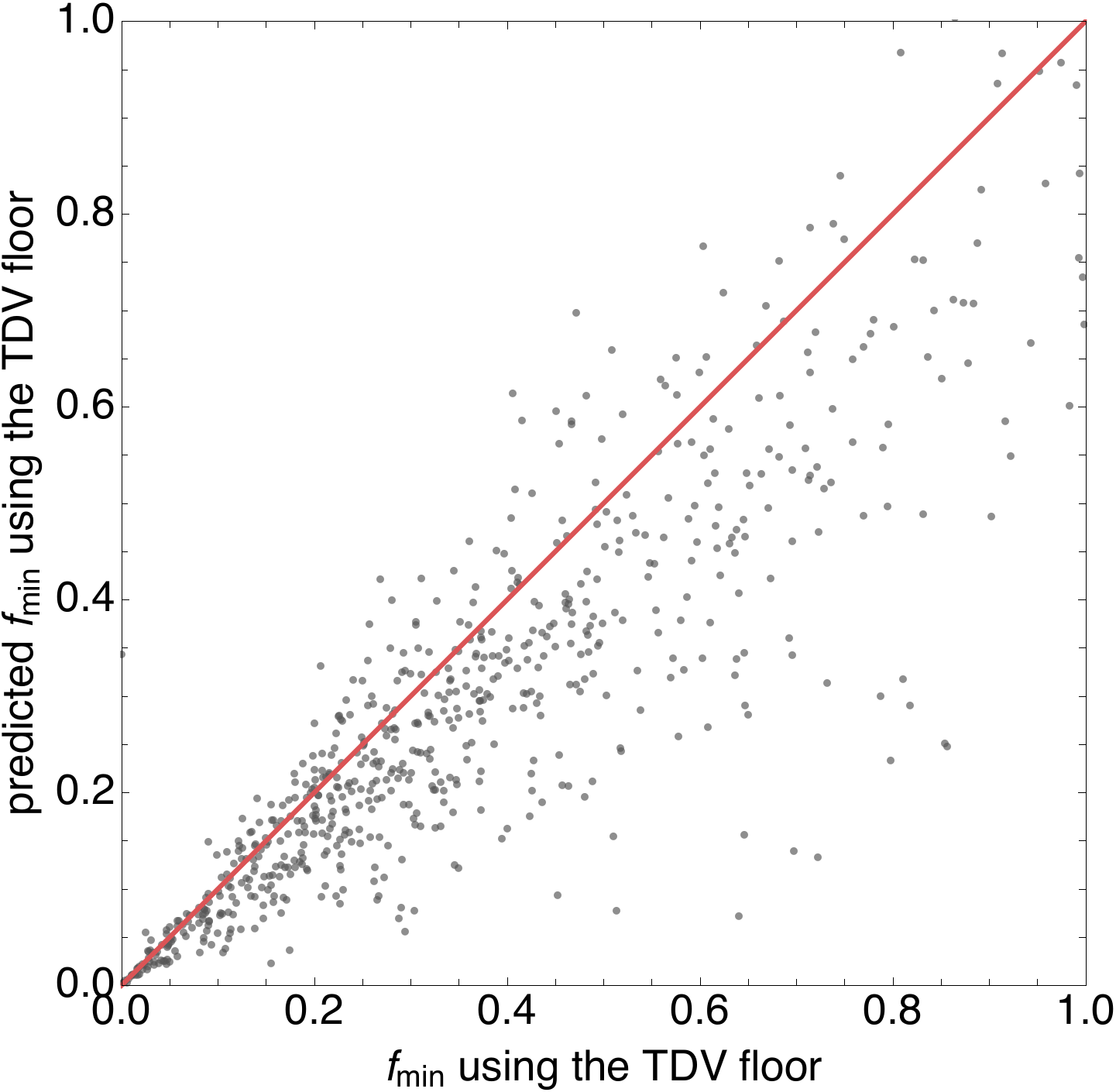}
\caption{
Comparison of the predicted lower limits on $f$ versus the actual obtained
values for 708 KOIs. It was found that the predictions are generally sound,
although there is some subset of KOIs for which the real TDV precision is
worse than the forecasted value.
}
\label{fig:comp}
\end{center}
\end{figure}

The accuracy of the predicted $f_{\mathrm{min}}$ values can be tested applying
the same procedure to the 708 KOIs in the full sample which truly do have TDVs.
In this way, one can plot the predicted $f_{\mathrm{min}}$ against the
calculated $f_{\mathrm{min}}$ values found earlier. This is shown in
Figure~\ref{fig:comp}, where a nearly 1:1 relation is obtained, as expected.
From the figure, one can see subset of KOIs for which the predicted
$f_{\mathrm{min}}$ are too optimistic, which is to be expected since the real
duration measurements can be occasionally degraded in precision due to effects
such as data gaps, star spots and flaring.

Finally, for the TTV ceiling, one needs an estimate of $q$ mass ratio between
the planet and the star. For this, the \citet{chen:2017} mass-radius relation
is used in what follows. In cases where the relation becomes degenerate
(specifically around a Jupiter radius), the upper allowed limit on the mass
is used, following the argument outlined earlier in
Section~\ref{sec:ttvceiling}. For 21 KOIs, the NEA does not list a planetary
radius (or transit depth) and thus it was not possible to estimate a TTV
ceiling for these cases\footnote{These are KOIs 2311.01, 2640.01, 4956.01,
5000.01, 5074.01, 5160.01, 5161.01, 5177.01, 5194.01, 5210.01, 5309.01,
5368.01, 5374.01, 5377.01, 5437.01, 5450.01, 5537.01, 5783.01, 5824.01,
5837.01, 5955.01.}.

The TTV ceiling criterion provides the greatest number of impossible moons, 179
KOIs. These consistently show the highest TTVs for planets of comparable
periods; the upper tail of the TTV amplitude distribution. These are surely
interesting TTV systems, being highly significant, but can be rejected for an
exomoon survey. These $f_{\mathrm{min}}$ limits are made available in
Table~\ref{tab}.

\begin{table*}
\caption{
Results from our application of the methods described in this work to the transit
timing catalog of the KOIs presented by \citet{holczer:2016}. Only a portion of the
table is shown here.
} 
\centering 
\begin{tabular}{c c c c c c c c} 
\hline\hline 
KOI &
$P$ [d] &
$A_{\mathrm{TTV}}$ [m] &
$\Delta(\mathrm{BIC})_{\mathrm{TTV}}$ &
$f_{\mathrm{min}}$ (TDV) &
$f_{\mathrm{min}}^{\mathrm{pred}}$ (TDV) &
$f_{\mathrm{min}}$ (TTV) &
TDVs? \\ [0.5ex] 
\hline 
1.01 & 2.471 & $0.0265 \pm 0.0082$ & 8.4 & 0.874 & 0.707 & 0.000154 & \checkmark \\
2.01 & 2.205 & $0.052 \pm 0.019$ & 2.8 & 1.44 & 1.45 & 0.000386 & \checkmark \\
3.01 & 4.888 & $0.133 \pm 0.033$ & 23.5 & 0.638 & 0.472 & 0.00404 & \checkmark \\
\dotfill & \dotfill & \dotfill & \dotfill & \dotfill & \dotfill & \dotfill & \dotfill \\[1ex]
\hline\hline 
\end{tabular}
\label{tab} 
\end{table*}

\subsection{Application to Kepler-1625b}

Additionally, the methods described in this work are applied to the only known
example of an exomoon candidate - Kepler-1625b i \citep{teachey:2018}.
The possible existence of Kepler-1625b i remains a topic of ongoing investigation.
For example, \citet{kreidberg:2019} recover the associated TTV signal but
not the moon-like dip, although \citet{looseends:2020} shows that their
analysis exhibits higher systematics. In contrast, \citet{heller:2019} recover
both the TTV and moon-like dip, but suggest an inclined hot-Jupiter as
an alternative hypothesis, which could be tested with precise radial
velocities. A detailed discussion of the candidacy of this object is beyond
the scope of this paper, but clearly applying our newly derived tests to the
only candidate transiting moon is a basic application expected for any new test.

Since no TDVs were explicitly derived in the original paper, it is first
necessary to obtain them. A starting point for this process are the transit
fits and processed data used by \citet{teachey:2018}. In particular, the ``T''
model  of that work is a suitable jumping off point, which assumes that no moon
is present and just fits the light curve trends and transit profile with a
simple \citet{mandel:2002} model. The ``T'' stands for TTVs, because the model
gives each epoch its own transit time as a free parameter, but the other
parameters, such as impact parameter and stellar density, are globally shared
across all epochs (which imposes a constant duration). Starting from this
model, it is updated to let $\rho_{\star}$ also be uniquely assigned to each
epoch. This increases the dimensionality of the ``T'' model by three new
parameters (four epochs minus the original global stellar density term). Since
the stellar density controls the planetary semi-major axis, which in turn
controls the planetary velocity, this inclusion allows for the measurement of
the TDV-Vs directly.

For the models that do not include a moon, the agreement between the different
trend models attempted by \citet{teachey:2018} are excellent and on this basis
the exponential long-term trend model is adopted in what follows.

With this new TDV model, the four light curves provided by \citet{teachey:2018}
are re-fit using \multi\ \citep{feroz:2008,feroz:2009}, yielding a
chronological sequence of durations (using definition $\tilde{T}$;
\citealt{investigations:2010}) of $1039_{-17}^{+18}$\,mins,
$1068_{-14}^{+16}$\,mins, $1034_{-68}^{+26}$\,mins and
$1063.7_{-5.4}^{+6.2}$\,mins. The most recent duration here comes from the HST
observations of \citet{teachey:2018}, which clearly provides substantial
improvement in precision.

The constant duration model performs well against these data, with $\chi^2 =
2.967$ for four data points. As an additional check on this, one may compare
the Bayesian evidences between the original ``T'' model of \citet{teachey:2018}
and this modified model which allows for TDVs. From this, it is found that the
original model is indeed favoured with a Bayes factor of $K = 87,000$ - a
decisive preference for a constant duration model. Accordingly, it is concluded
that there exists no evidence for detectable TDVs. This is perhaps not
surprising given that the \citet{teachey:2018} solution places the moon at a
fairly wide separation, where TDVs are attenuated.

Having established that no TDVs exist, the next step is to calculate an upper
limit on the TDV amplitude. This is straight-forwardly achieved by fitting an
offset + sinusoid to the derived durations using a simple MCMC, infers a
3\,$\sigma$ upper limit on the TDV amplitude of 55\,mins.

It is now possible to use the TDV upper limit along with the TTV amplitude
to infer the minimum allowed exomoon semi-major axis, without ever
formally fitting the data to an exomoon model. The TTV amplitude is
not actually fit per say in \citet{teachey:2018}, but rather is
incorporated into their photodynamical planet+moon solution. From
the reported TTVs, a sinusoid was regressed to obtain
$A_{\mathrm{TTV}} = (19.1 \pm 1.9)$\,minutes. The ratio of the
quoted uncertainty compared to the amplitude indicates that this is
highly significant and indeed this has already been established
through the Bayes factor comparison in \citet{teachey:2018}, who
find a $K = 10.0$ in favor of the TTV model over a static case.
Plugging these numbers into Equation~(\ref{eqn:fmin}) yields
$f > 0.046$.

For the TTV ceiling effect, a rough estimate can
be found by simply assuming the planet is of order of a Jupiter mass,
given that it approximately a Jupiter radius. Using
Equation~(\ref{eqn:fmin2}), this yields $f > 0.0042$. Thus, in this
case, the TDV floor limit imposes a tighter constraint. Certainly
then, there is plenty of room for an exomoon given the current
timing measurements. Indeed, the exomoon solution of
\citet{teachey:2018} places the moon at $f\simeq0.2$, which satisfies
this minimum constraint.

\section{Discussion}

This work derives two distinct, but related, means of calculating a lower limit
on an exomoon's semi-major axis divided by its planetary Hill radius,
conditioned upon the hypothesis that an observed (and significant) TTV is
solely caused by a single exomoon. If either of these lower limits exceeds
unity (i.e. the moon is outside the Hill sphere), the hypothesis can be
rejected on the via an argument of \textit{reductio ad absurdum}. One is free
to modify the critical limit to less than unity, under more conservative
assumptions regarding the range of stable moons (e.g. see
\citealt{domingos:2006}).

The two methods both assume that a TTV detection has been made and that one
has in hand a TTV amplitude. They differ in whether they assume an upper limit
on the TDV amplitude has been computed. Most directly, our work asks whether
the detected TTV amplitude could plausibly be caused by an exomoon? Of course,
if a signal passes these tests, that does not mean that one can necessarily claim
an exomoon detection on this basis alone - follow-up investigations
using tools such as photodynamics will be needed (e.g. \citealt{teachey:2018}).
However, failing one of the criteria is damning for the hypothesis that the
observed TTVs are solely caused by a single exomoon. Nevertheless, one could
suggest that some component of the TTVs are still caused by one (or more)
moons, with other TTVs effects contributing to the total - although the
additional complexity of such a hypothesis would certainly necessitate some
specific motivation to justify.

The two limits are given by Equations~(\ref{eqn:fmin}) \& (\ref{eqn:fmin2}).
Since both require the TTV amplitude, but only the former requires the TDV
upper limit, it is worthwhile pausing to ask whether the former ever
realistically undercuts the latter, or can we generally rely on
Equation~(\ref{eqn:fmin2}) alone for the best limits? From the applied
examples in this work, using the \citet{holczer:2016} \kepler\ transit timing
catalog, 40 impossible moons were identified via Equation~(\ref{eqn:fmin}),
and 179 via Equation~(\ref{eqn:fmin2}), but critically only one object which
appears in both lists (KOI-5497.01). Tightening the $f$ constraint to only
prograde moons ($f<0.5$) yields many more impossible moons for both methods
(see Figures~\ref{fig:tdv} \& \ref{fig:ttv}) yet only one new overlapping
object (KOI-4927.01). It is further highlighted that the TTV ceiling requires
an estimate of the planet-to-star mass ratio, whereas the TDV floor does not.
Thus, the two limits are highly complementary and the use of both is
recommended.

This work highlights that there are a considerable number of KOIs for
which TDVs would be useful. Of the 1709 KOIs without TDVs in the used sample,
it was found that 22 should be expected to yield a TDV upper limit sufficient
to completely exclude moons (since $f_{\mathrm{min}}>1$).
Of these 22, 11 already have a TTV ceiling constraint that indicates an
impossible moon. For the other 11 (KOIs 72.01, 452.01, 732.01, 1300.01,
1428.01, 2215.01, 2573.01, 3913.01, 4927.01, 5128.01, 5713.01), the current
TTV constraints do not fully prohibit an exomoon and thus these would make
interesting objects for follow-up investigations.

\begin{figure}
\begin{center}
\includegraphics[width=8.4cm]{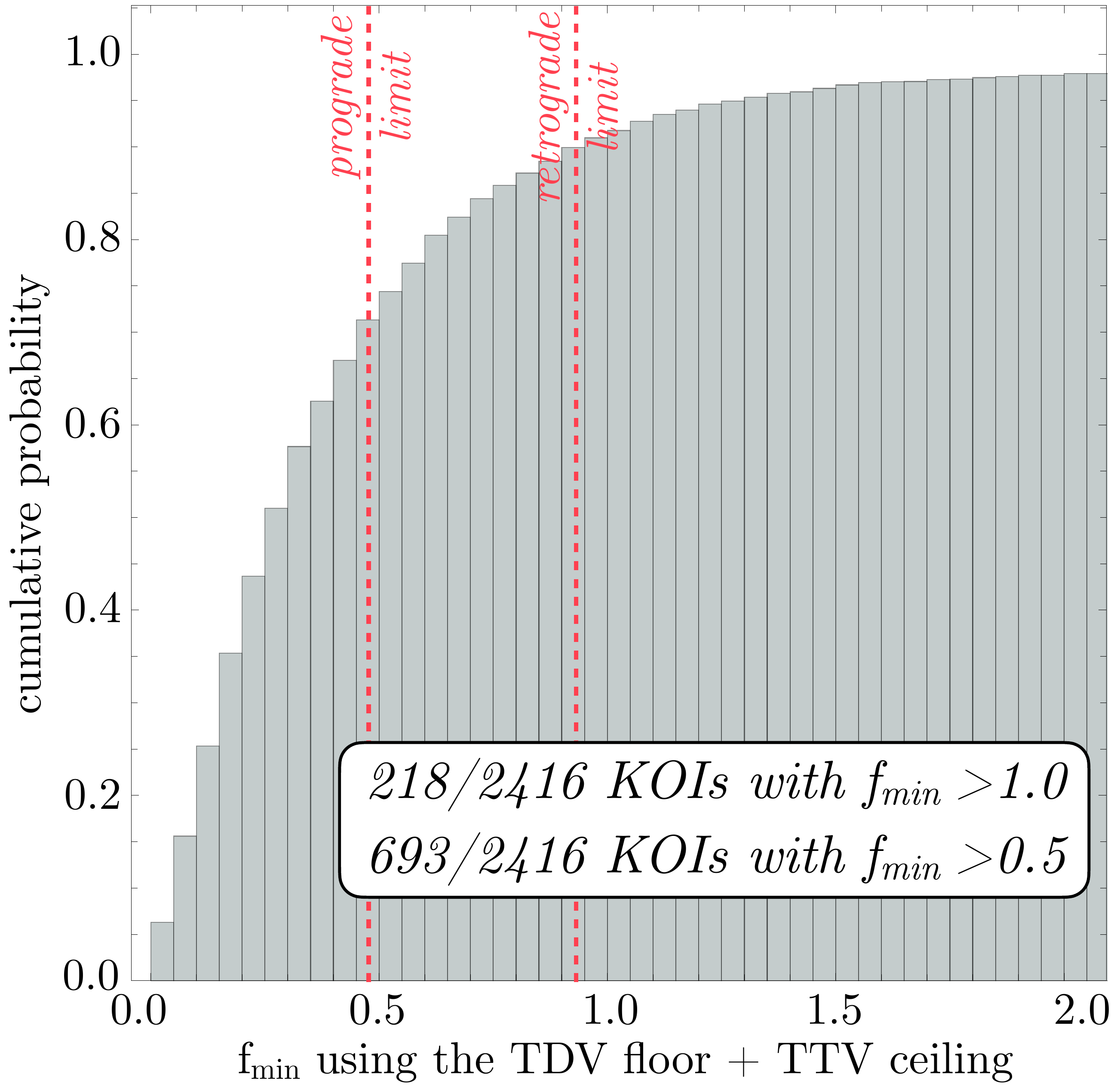}
\caption{
Cumulative histogram of the $f_{\mathrm{min}}$ value derived using both
the TDV floor and TTV ceiling together (taking the maximum of the two)
for the KOIs in our sample. About 9\% of the sample can be rejected as being
plausibly due to an exomoon using the most generous assumptions about
moon stability ($f<1$), and 29\% using $f<0.5$.
}
\label{fig:cdf}
\end{center}
\end{figure}

Repeating this for a critical $f$ threshold of 0.5 increases the number of
KOIs deserving of TDV follow-up to 109 - Table~\ref{tab} provides a full list
of these. A cumulative histogram of the best possible constraint on $f$ is
shown Figure~\ref{fig:cdf} for the full sample used in this work.
 
It is briefly highlighted that application of these techniques to Kepler-1625b
offers another example of a test that this moon candidate survives. As
emphasized earlier, passing this test does not itself prove the case for a
moon, but failure to do would have provided a simple means to discard the
exomoon candidate.

As demonstrated from these examples, the tests discussed here provide a simple
and well-motivated test to remove spurious signals in the search for exomoons.
Their application is encouraged to those looking for such effects, as an
expedient means of removing false-positives. More broadly, this work showcases
the value of timing effects and the benefits of community derived TTVs and
TDVs.


\acknowledgements{
DK is supported by the Alfred P. Sloan Foundation.
AT is supported through the NSF Graduate Research Fellowship (DGE-1644869).
Thanks to Dan Fabrycky for helpful comments on a preliminary draft.
Special thanks to Tom Widdowson, Mark Sloan, Laura Sanborn, Douglas Daughaday, Andrew Jones, Jason Allen, Marc Lijoi, Elena West, Tristan Zajonc, Chuck Wolfred, Lasse Skov, Martin Kroebel and Geoff Suter.
This work is based in part on observations made with the NASA/ESA Hubble Space Telescope, obtained at the Space Telescope Science Institute, which is operated by the Association of Universities for Research in Astronomy, Inc., under NASA contract NAS 5-26555. These observations are associated with program \#GO-15149. Support for program \#GO-15149 was provided by NASA through a grant from the Space Telescope Science Institute, which is operated by the Association of Universities for Research in Astronomy, Inc., under NASA contract NAS 5-26555. This paper includes data collected by the Kepler mission. Funding for the Kepler mission is provided by the NASA Science Mission directorate.
This research has made use of the NASA Exoplanet Archive, which is operated by the California Institute of Technology, under contract with the National Aeronautics and Space Administration under the Exoplanet Exploration Program.
}


\vskip2mm

\newcommand\eprint{in press }

\bibsep=0pt

\bibliographystyle{aa_url_saj}

{\small

\bibliography{sample_saj}
}

\begin{strip}

\clearpage

{\ }

\clearpage

{\ }

\newpage

{\ }



\naslov{UPU{T}{S}TVO ZA AUTORE}


\authors{B. Arbutina$^{1}$, D. Uro{\v s}evi{\' c}$^{1}$ and M. Jovanovi{\' c}$^2$}

\vskip3mm


\address{$^1$Department of Astronomy, Faculty of Mathematics,
University of Belgrade\break Studentski trg 16, 11000 Belgrade,
Serbia}


\Email{arbo@math.rs, dejanu@math.rs}

\address{$^2$Astronomical Observatory, Volgina 7, 11060 Belgrade 38, Serbia}

\Email{milena@aob.rs}

\vskip3mm


\centerline{{\rrm UDK} \udc}


\vskip1mm

\centerline{\rit Uredjivaqki prilog}

\vskip.7cm




\begin{multicols}{2}

{
\rrm

Ovaj prilog ima za cilj da pomogne autorima kod pripreme
qlanaka za na{\ss} qasopis.~Radove, po mogu{\cc}nosti, treba
pripremiti koriste{\cc}i}  \LaTeX\ {\rrm uz posebnu datoteku koja
defini{\ss}e izgled qasopisa} Serbian Astronomical Journal {\rrm i
sadr{\zz}i unapred definisana polja u preambuli dokumenta i neke
dodatne ili de\-li\-miq\-no izmenjene komande koje se koriste pri unosu
samog teksta.~Autori mogu koristiti ovaj dokument kao primer pri
kucanju svojih radova.

{\ }

}

\end{multicols}

\end{strip}


\end{document}